\documentclass[10pt,conference]{IEEEtran}

\usepackage[T1]{fontenc}
\usepackage{graphicx}
\usepackage{amsmath}
\usepackage{amssymb}
\usepackage{cite}
\usepackage{url}
\usepackage{microtype}
\usepackage{booktabs}
\usepackage{balance}
\usepackage{capt-of}
\usepackage{xcolor}
\usepackage{newtxtext,newtxmath}
\newcommand{\best}[1]{\textbf{#1}}
\newcommand{\na}{\multicolumn{1}{c}{--}}
\newcommand{\desctext}{\small}
\definecolor{linkteal}{rgb}{0.15,0.45,0.55}
\usepackage[colorlinks=true,
            urlcolor=linkteal,
            linkcolor=black,
            citecolor=black]{hyperref}
\makeatletter
\newcommand\blfootnote[1]{\begingroup\renewcommand\thefootnote{}\footnote{#1}\addtocounter{footnote}{-1}\endgroup}
\makeatother

\title{Graph Reinforcement Learning for Calibration-Aware Quantum Circuit Routing}

\author{
\IEEEauthorblockN{Yash Vardhan Tomar}
\IEEEauthorblockA{\textit{Purdue University}\\[-1pt]
West Lafayette, USA\\[-1pt]
tomar4@purdue.edu}
\and
\IEEEauthorblockN{Dheeraj Peddireddy}
\IEEEauthorblockA{\textit{Purdue University}\\[-1pt]
West Lafayette, USA\\[-1pt]
dpeddire@purdue.edu}
}

\begin{document}
\raggedbottom
\maketitle

\blfootnote{\copyright~2026 IEEE. Personal use of this material is permitted. Permission from IEEE must be obtained for all other uses, in any current or future media, including reprinting/republishing this material for advertising or promotional purposes, creating new collective works, for resale or redistribution to servers or lists, or reuse of any copyrighted component of this work in other works.}

\begin{abstract}
Quantum circuit routing is a key step in compiling programs for noisy intermediate-scale quantum processors, particularly superconducting devices whose sparse fixed coupling makes routing a central compilation cost. Routes that appear efficient by standard overhead metrics such as SWAP count, routed two-qubit count, and depth can still lose fidelity when they pass through poorly calibrated couplers. We study a calibration-aware graph reinforcement-learning router that uses same-day calibration data from superconducting IBM Heron r2 processors to choose hardware-edge SWAPs. We train the policy with proximal policy optimization and evaluate it with exact simulated fidelity across nine Munich Quantum Toolkit (MQT) Bench circuits and three calibration snapshots. Across these evaluations, pooled mean exact fidelity is $0.727$, compared with $0.440$ for SWAP-based bidirectional heuristic search (SABRE)-best20 and $0.481$ for target-aware SABRE. We observe that fidelity gains come with higher routed two-qubit counts and are concentrated in 5 qubit and 8 qubit circuit families; under the fixed tree action graph, all 10 qubit families favor SABRE-best20. Overall, our results show that calibration-aware learned routing can improve fidelity beyond gate-count-driven compilation, by roughly $0.25$ to $0.29$ in absolute mean fidelity over the SABRE-family baselines.
\end{abstract}

\par\vspace{2pt}


\begin{figure*}[!t]
\centering
\noindent\hspace*{\dimexpr-1in-\oddsidemargin\relax}\makebox[\paperwidth][c]{\includegraphics[width=\paperwidth,trim=0 26pt 0 0,clip]{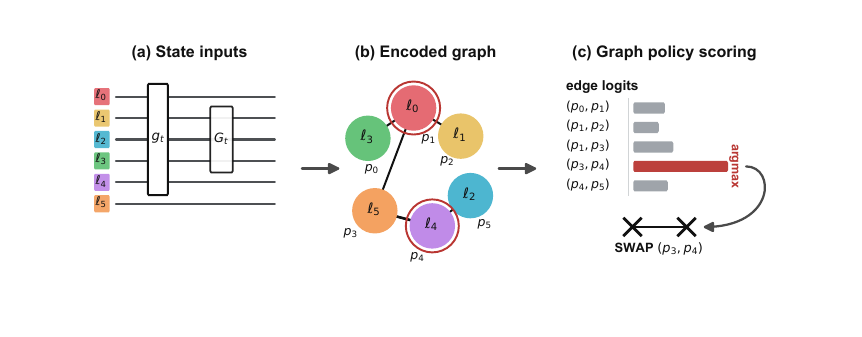}}
\refstepcounter{figure}\label{fig:method-overview}
\par\vspace{0pt}
\noindent\parbox{0.95\textwidth}{\raggedright\desctext\textbf{Fig. \thefigure.} Routing-state graph construction. \textbf{(a)} Circuit inputs consist of the remaining circuit, front blocking gate $g_t$, and lookahead gates $G_t$. \textbf{(b)} Current non-identity placement $M_t:L\rightarrow P$ and calibration snapshot $\kappa$ are encoded on $G_B=(P,E,\kappa)$; node labels show logical occupants, and red rings mark the blocked front-gate operands. A solid $p_3$--$p_4$ edge marks a legal SWAP edge on the shortest path. \textbf{(c)} Message passing produces node embeddings, the edge head scores legal SWAPs, and the red logit identifies the inserted SWAP.}
\end{figure*}

\section{Introduction}
Near-term quantum processors, known as noisy intermediate-scale quantum (NISQ) devices, have reached scales capable of executing nontrivial circuits without the safety net of full error correction~\cite{preskill2018nisq}.
To successfully test quantum algorithms that challenge classical capabilities, their compiled execution must rigorously defend the intended state against pervasive hardware noise.
Compilation must turn an ideal logical circuit into a device-level program while respecting sparse connectivity, finite coherence, and calibration data that varies across qubits, couplers, and calibration cycles.
Logical circuits may request two-qubit interactions between any pair of logical qubits, while the processor can execute them only along calibrated physical couplers.
We focus on superconducting processors, where routing is most consequential because the fixed, sparse coupling map and per-coupler calibration force most nonlocal interactions through inserted SWAPs.

To execute those interactions, a compiler must route the circuit through a hardware graph by selecting a physical placement and inserting SWAP operations whenever a requested interaction involves nonadjacent physical qubits. Routing is often summarized by SWAP count, routed two-qubit count, or depth, which makes minimizing operations seem like the natural objective. On calibrated hardware, that shortcut can fail because two routes with nearly the same overhead may traverse couplers with very different error rates, and a route with more gates can sometimes preserve more of the ideal state. 

Classical routing work has long associated SWAP minimization with token-swapping and NP-hard mapping formulations~\cite{siraichi2019allocation,cowtan2019routing}, and production compilers rely on fast heuristics such as SWAP-based bidirectional heuristic search (SABRE) and backend-aware Qiskit passes~\cite{zulehner2019mapping,li2019sabre,qiskit}. Calibration-aware compilation has demonstrated that backend error data can influence which placement or route is optimal for a circuit~\cite{murali2019noise,nishio2020erroraware,tannu2019ensemble,pascoal2024noise}. Learned routers, including AlphaRouter, approach routing as a sequential decision problem where the policy selects SWAPs based on the circuit state~\cite{tang2024alpharouter}.

These results lead to the comparison we test here, where SABRE-style heuristics provide strong, reproducible baselines, calibration-aware compilers demonstrate that coupler error rates matter, and the learned policy must use same-day calibration while still observing the remaining circuit and current placement. This setup raises the question: \emph{Can a learned router use same-day calibration data to improve exact simulated state fidelity over reproducible SABRE-family baselines, even when the route uses more two-qubit gates?}

We evaluate a SABRE-anchored calibration-aware proximal policy optimization (PPO) router on nine Munich Quantum Toolkit (MQT) Bench circuits and three same-day IBM Heron r2 calibration snapshots~\cite{niemann2023mqtbench}. Figure~\ref{fig:method-overview} illustrates how the remaining circuit, non-identity placement, and calibration snapshot are encoded on the hardware graph before legal SWAP edges are scored.

We observe fidelity gains mainly on the 5q and 8q circuit families, where additional two-qubit gates can steer the circuit away from less reliable calibrated couplers.
The fixed tree action graph leaves too few alternatives for the 10q families, where SABRE-best20 remains better, and we treat this reversal as central to the evaluation because calibration-aware routing can help only when the action graph offers useful alternatives.

\section{Related Work}
Noise-adaptive mappings, error-aware compilation, and ensemble mapping methods find that backend heterogeneity can change which placement is best for a circuit~\cite{murali2019noise,nishio2020erroraware,tannu2019ensemble}.
Recent router benchmarks and instruction-set-architecture-aware compilers issue the same caution empirically, showing that gate-count-only comparisons can hide routes that behave differently once the hardware error model is included~\cite{pinacanelles2025routers,yang2025canopus}.

Learning-based routing builds on reinforcement learning (RL)~\cite{schulman2017ppo} and graph neural networks (GNNs) for circuit and hardware states~\cite{gilmer2017mpnn,kipf2017gcn}.
Prior examples include deep-Q-network (DQN) routing, GNN-aided Monte Carlo tree search (MCTS), PPO with IBM calibration data, and tree-search RL in AlphaRouter~\cite{pozzi2020routing,sinha2021gnn,pascoal2024noise,tang2024alpharouter}.
Among these learned-router studies, Pascoal et al.~\cite{pascoal2024noise} is the closest methodological reference, while our evaluation moves the policy input onto the hardware graph, reports exact density-matrix state fidelity, and uses same-day Heron r2 snapshots.
Because published checkpoints, Heron-compatible evaluation harnesses, and action spaces differ from our protocol, the experiments compare against reproducible SABRE-family baselines and leave a matched learned-router reimplementation for an expanded study.

\section{Method and Benchmark Protocol}

We model routing as a finite sequence of legal hardware-edge SWAP choices on a calibrated graph, so calibration enters at the same level as the action set and the policy sees hardware quality while scoring legal SWAPs.
Let $G_B=(P,E,\kappa)$ be the calibrated backend graph, with selected physical qubits $P$, executable couplers $E$, and snapshot calibration data $\kappa$ containing readout error, one- and two-qubit error, and coherence.
At step $t$, the remaining logical program $C_t=(L,G_t,\prec)$ contains logical qubits $L$, remaining gates $G_t$, and program order $\prec$.
Placement $M_t:L\rightarrow P$ maps each logical qubit to its physical location, giving routing state $s_t=(C_t,M_t,\kappa)$.
We write $g_t$ for the first nonexecutable two-qubit gate in the ordered remainder, and its operands form the blocking pair used by the observation.
Choosing action $a_t=(p_i,p_j)\in E$ inserts a SWAP on a hardware edge and exchanges the logical occupants of $p_i$ and $p_j$.
After each SWAP, the environment greedily appends the longest executable prefix; one-qubit gates are always executable, a two-qubit gate with logical operands $(u,v)$ is executable exactly when $\{M_t(u),M_t(v)\}\in E$, and an episode ends after the full program has been appended or the timeout is reached.

For $Q$ physical qubits, each observation has dimension $6Q+4$, with node features for readout error, coherence, incident two-qubit error, hosted logical index, lookahead demand distance, and next-gate flag.
Four global features encode the blocking-pair endpoints, their shortest-path distance, and routed progress; demand distance summarizes how far a hosted logical qubit is from its partners in the remaining gate window, and the next-gate flag marks the operands of $g_t$.
Let $x_i$ denote the node feature vector at physical qubit $p_i$, and let $e_{ij}$ denote edge attributes containing calibrated two-qubit error and the legal-action mask.
Given this graph-structured observation, the policy runs two message-passing layers on $G_B$ to produce node embeddings $h_i$.
We score each legal edge with a multilayer perceptron over $(h_i,h_j,e_{ij})$, and a masked softmax defines $\pi_\theta(a_t\mid s_t)$, where $\theta$ denotes the trainable policy parameters.
Because routing has long horizons and sparse terminal feedback, we train $\pi_\theta(a_t\mid s_t)$ with PPO~\cite{schulman2017ppo}, generalized advantage estimation (GAE), shaped intermediate rewards, and the settings in Table~\ref{tab:model-hparams}.
Here $\gamma$ is the reward discount, GAE $\lambda$ is the advantage-trace parameter, and ``clip'' is the PPO probability-ratio clipping threshold.

\par\vspace{8pt}
\noindent\begin{minipage}{\columnwidth}
\centering
\refstepcounter{table}\label{tab:model-hparams}
{\small
\setlength{\tabcolsep}{0pt}
\renewcommand{\arraystretch}{1.03}
\begin{tabular*}{0.86\columnwidth}{@{\extracolsep{\fill}}lr@{}}
\toprule
\multicolumn{2}{@{}l}{\textit{Graph policy}} \\
Message-passing layers & 2 \\
Hidden dimension & 64 \\
Node-feature dimension & 10 \\
Observation dimension & $6Q+4$ \\
Trainable parameters & 67k \\
\midrule
\multicolumn{2}{@{}l}{\textit{PPO optimization}} \\
Optimizer & AdamW \\
Learning rate & $3.0{\times}10^{-4}$ \\
PPO epochs & 4 \\
Batch size & 128 transitions \\
$\gamma$, GAE $\lambda$, clip & 0.98, 0.99, 0.28 \\
Entropy coefficient / floor & 0.37 / 0.02 \\
\bottomrule
\end{tabular*}}
\par\vspace{3pt}
\noindent\parbox{0.97\columnwidth}{\raggedright\desctext\textbf{Table \thetable.} Graph-policy architecture and PPO settings.}
\end{minipage}
\par\vspace{7pt}

\begin{figure*}[!t]
\centering
\includegraphics[width=0.8\textwidth]{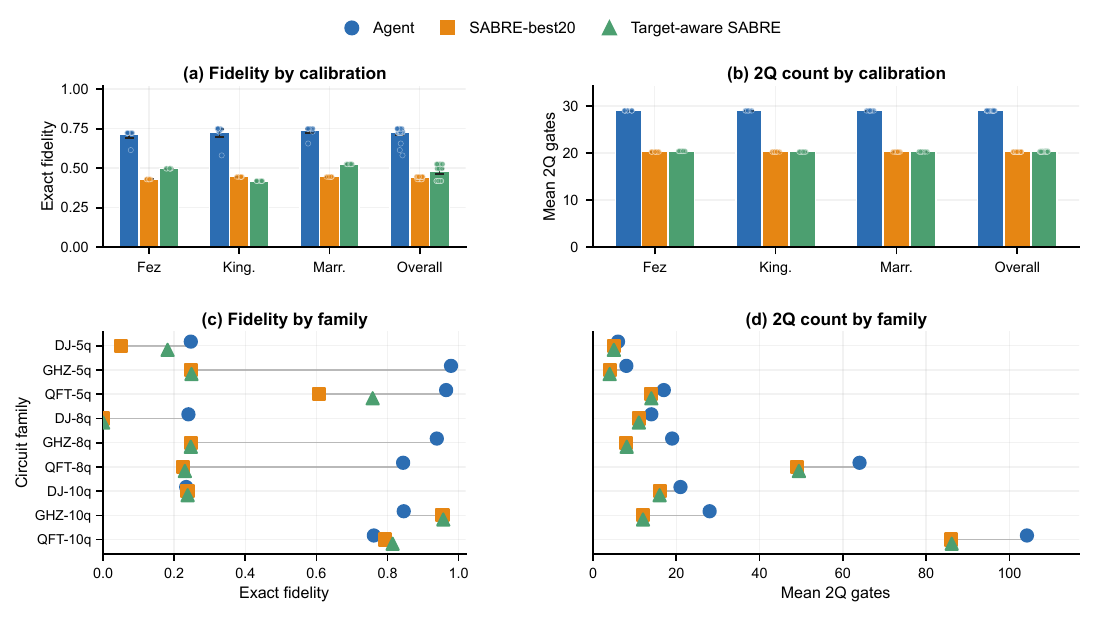}
\refstepcounter{figure}\label{fig:benchmark-results}
\par\vspace{0pt}
\noindent\parbox{1.00\textwidth}{\raggedright\desctext\textbf{Fig. \thefigure.} Benchmark exact fidelity and routed two-qubit (2Q) counts. (a,b) Snapshot means with matched cells and bootstrap intervals. (c,d) Circuit-family means. Fidelity gains coincide with additional 2Q gates on families where calibration-aware routing helps.}
\end{figure*}

Exact density-matrix fidelity is the evaluation target, but recomputing it after every PPO action would dominate rollout cost, so training uses a low-cost calibration-aligned proxy and reserves exact simulator fidelity for final evaluation.
For a nonterminal transition,
\begin{equation}
r_t=\lambda_d\Delta d_t+\lambda_\rho\Delta\rho_t+\lambda_g n_t-\lambda_s-\lambda_i\mathbf{1}_{\rm invalid}.
\end{equation}
Here $\Delta d_t$ is the reduction in normalized shortest-path distance for the next blocking two-qubit gate, $\rho_t$ is routed-program progress, $\Delta\rho_t$ is its positive change, $n_t$ is the number of logical gates appended after the SWAP, $\lambda_s$ charges each valid SWAP, and $\mathbf{1}_{\rm invalid}$ equals one exactly when the chosen edge is illegal under the current action mask.
At route completion, the terminal reward adds
\begin{equation}
r_T \leftarrow r_T+\lambda_f(\hat F_A-\hat F_S)-\lambda_c(J_A-J_S),
\end{equation}
where $\hat F_A$ and $\hat F_S$ are agent and SABRE-best20 proxy fidelities for the same circuit and snapshot, while $J_A$ and $J_S$ are route-cost scores computed as $J=g_{2Q}+0.01d$ for the corresponding routed circuits.
Here $g_{2Q}$ is routed two-qubit (2Q) count, $d$ is routed depth, and the $0.01$ coefficient makes 2Q count primary while using depth as a tie-breaker.
Timeouts subtract 0.5, and unfinished episodes subtract $10(1-\rho_T)$, where $\rho_T$ is the final routed-program progress at termination.
We use fixed weights $\lambda_d=0.05$, $\lambda_\rho=2.0$, $\lambda_g=0.01$, $\lambda_s=0.02$, $\lambda_i=0.2$, $\lambda_f=10.0$, and $\lambda_c=0.01$.
The proxy is estimated success probability (ESP)~\cite{nishio2020erroraware} over the routed circuit,
\begin{equation}
\hat F_{\rm raw}(R)=\prod_{g\in R}(1-p_g),\quad
\hat F=\bigl(1-\mathrm{clip}(k(1-\hat F_{\rm raw}^{1/|R|}),0,1)\bigr)^{|R|},
\end{equation}
\par\vspace{2pt}
for routed circuit $R$.
Here $p_g$ is the calibrated error probability assigned to gate $g$, $|R|$ is the routed-circuit gate count, and $\mathrm{clip}(x,0,1)$ truncates $x$ to the interval $[0,1]$.
We set $k=1$ to keep the proxy on the raw ESP scale and avoid introducing an additional fitted rescaling constant.

Final evaluation uses the same calibration data for environment construction and simulation, with backend snapshots defining the topology and parameterizing the noisy density-matrix simulator; exact fidelity is reported as $F=\langle\psi|\rho|\psi\rangle$, with simulated output $\rho$ and ideal target state $|\psi\rangle$.
Our Qiskit Aer model uses symmetric readout error, averaged one-qubit depolarizing error, calibrated two-qubit depolarizing error clipped to 0.30, a single-coupling SWAP depolarizing channel, and thermal relaxation with a clamped $\min(T_1,T_2)$ coherence proxy in $[1,300]\,\mu\mathrm{s}$; real Heron SWAPs decompose to three controlled-X (CX) gates.

Each episode jitters readout/coherence by 10\% and one-/two-qubit error by 15\%, shared across paired methods, so the simulated calibrations vary modestly while the paired comparisons remain fixed.

Our benchmark uses nine fixed pre-compilation MQT Bench Open Quantum Assembly Language (QASM) circuits~\cite{niemann2023mqtbench}, run through Qiskit~\cite{qiskit}, with three families at 5, 8, and 10 qubits.

Same-day Heron r2 snapshots \emph{Fez}, \emph{Kingston}, and \emph{Marrakesh} begin as 156-qubit targets, from which we use one connected 10-qubit tree action graph with nine calibrated couplers; topology remains fixed across experiments, while calibration values vary across snapshots.
Each snapshot has 10 seed-specific evaluation conditions with 50 episode pairs per condition, for 1500 paired episodes in total.
Targets are sampled uniformly from the nine compatible QASM circuits, and ten PPO runs are evaluated after 40,000 training episodes each to obtain matched condition means while keeping exact density-matrix evaluation tractable.

Both reproducible SABRE-family baselines~\cite{li2019sabre} are selected over the same 20-seed sweep: SABRE-best20 selects the route minimizing the routing cost $J=g_{2Q}+0.01d$ defined above, while target-aware SABRE uses a calibration-derived Qiskit \texttt{Target} with SABRE layout/routing and optimization level 3, then selects first by ESP proxy fidelity and then by the same cost score.

\section{Results and Discussion}

In the fixed in-distribution evaluation, the learned policy improves exact fidelity over both SABRE-family baselines.
Across all matched conditions, pooled mean exact fidelity is 0.7269 for the learned policy, compared with 0.4402 for SABRE-best20 and 0.4807 for target-aware SABRE.
All three snapshot means and all 30 condition means favor the learned router; after grouping by circuit size, however, every 10q family favors SABRE-best20 on the fixed tree action graph.

Because the fidelity gains come with measurable overhead, Fig.~\ref{fig:benchmark-results} reports routed two-qubit counts alongside fidelity; relative to SABRE-best20, the learned routes add +8.63 two-qubit gates and +4.61 depth.
Using routed two-qubit count $g_{2Q}$, $\bar{\epsilon}_{\mathrm{eff}}=-\ln(\bar F)/\bar g_{2Q}$ gives 0.011 for the agent and 0.041/0.036 for the baselines.
Framed as a cost--benefit question, whether the fidelity gain justifies the added two-qubit gates depends on where the extra gates are placed: pooled over all conditions, the learned router raises exact fidelity from 0.440 to 0.727, a 0.287 absolute and roughly 65\% relative gain over SABRE-best20, in exchange for 8.63 additional two-qubit gates. Because those gates are routed onto better-calibrated couplers, the effective per-2Q-gate error $\bar{\epsilon}_{\mathrm{eff}}$ falls by nearly a factor of four, so the longer routes are net favorable rather than simply more expensive. This favorable balance holds on the 5q and 8q families, where avoiding unreliable couplers repays the added gates; on the 10q families, the fixed tree action graph forces the same overhead without a compensating fidelity gain, and the two-qubit increase is not justified.
Across all 30 conditions, the highest-fidelity routes also have larger gate counts, so gate count remains useful for overhead but incomplete as a fidelity proxy on calibrated hardware.

\par\vspace{12pt}
\noindent\begin{minipage}{\columnwidth}
\centering
\refstepcounter{table}\label{tab:paired-effects}
{\small
\setlength{\tabcolsep}{0pt}
\renewcommand{\arraystretch}{1.08}
\begin{tabular*}{\columnwidth}{@{\extracolsep{\fill}}lcc@{}}
\toprule
Comparison & Mean $\Delta$ [95\% interval] & $p$ \\
\midrule
Fidelity vs.\ SABRE-best20 & \best{+0.2867 [0.2712, 0.2988]} & $1.5{\times}10^{-6}$ \\
Fidelity vs.\ Target-aware & \best{+0.2461 [0.2250, 0.2672]} & $1.5{\times}10^{-6}$ \\
\addlinespace[2pt]
Extra 2Q vs.\ SABRE-best20 & +8.63 & \na \\
Extra depth vs.\ SABRE-best20 & +4.61 & \na \\
\bottomrule
\end{tabular*}}
\par\vspace{3pt}
\noindent\parbox{\columnwidth}{\raggedright\desctext\textbf{Table \thetable.} Paired effects over 30 snapshot-seed conditions. Wilcoxon signed-rank tests use matched condition means.}
\end{minipage}
\par\vspace{12pt}

\section{Conclusion}
In this work, we studied whether same-day calibration data can help a learned policy choose higher-fidelity SWAPs, even when those choices add gates, by providing the PPO router with the remaining circuit, current placement, and calibrated IBM Heron r2 hardware graph before it scores legal hardware-edge SWAPs. Across nine MQT Bench circuits and three calibration snapshots, this graph policy achieves a pooled mean exact fidelity of $0.727$, compared with $0.440$ for SABRE-best20 and $0.481$ for target-aware SABRE.

These fidelity gains commonly appeared in 5q and 8q families, where additional two-qubit gates could avoid less reliable couplers. In the 10q families, the fixed tree action graph offers too few alternate paths, and SABRE-best20 performs better. The paired gains and reversals show that calibration can justify longer routes when the hardware graph provides the router with useful alternatives, while a narrow action graph can turn extra gates into overhead.
Future work should evaluate cyclic Heron subgraphs, held-out circuits, hardware runs, and matched learned-router baselines; routing comparisons should report fidelity and calibration context alongside gate-count overhead.

\par\vspace{4pt}

\noindent\textbf{Data and code availability statement.} The depicted data and the used code can be found at \url{https://github.com/YTomar79/calibration-aware-rl-routing}.

\end{document}